\documentstyle[version2,aps]{revtex}
\begin{document}
\draft
\columnsep -.375in
\twocolumn[
\begin{title}
Interactions and Disorder in Multi-Channel Quantum Wires
\end{title}
\author{
N.\ P.\ Sandler $^{(a)}$ and  Dmitrii  L.\ Maslov $^{(a,b,c)}$}
\begin{instit}
$^{(a)}$Department of Physics and
$^{(b)}$Materials  Research Laboratory\\
University of Illinois at Urbana-Champaign,
Urbana, Illinois  61801, USA;\\
$^{(c)}$Department of Physics\\
University of Florida, P.O. Box 118440, Gainesville, Florida 32611
\cite{Maslov}\\
\end{instit}
\receipt{\today}
\begin{abstract}
Recent experiments have revealed that the temperature dependence of the
conductance of quasi-ballistic quantum wires bears clear features
of the Luttinger-liquid state. In this paper, the conductance of an
$N$-channel quantum wire is calculated within the model of N coupled
Luttinger liquids and under the assumption of weak disorder. It is
shown that as the number of channels increases, a crossover from the
Luttinger-liquid to the Fermi-liquid behavior occurs. This crossover
manifests itself in the $1/N$ decrease of the scaling exponent of 
the temperature dependence. An exact expression for the scaling
exponent for the case of N coupled Luttinger chains is obtained,
and the large $N$ limit is studied for the case of a quantum wire.
The case of $N=2$ for electrons with spin is analyzed in detail,
and a qualitative agreement with the experiment is achieved.
\end{abstract}
]

\narrowtext
\section{INTRODUCTION}
\label{sec:intro}

Although the conventional Landau theory for Fermi liquids \cite{ REF:Baym, REF:Pines} has been very successful in understanding  
many condensed matter systems, its failure to describe quasi-one-dimensional (Q1D) systems, {\it e.g.}, 
conducting polymers and organic conductors,  has
 motivated the search for alternative models which can describe a non-Fermi liquid 
behavior. 
The most theoretically studied example of a non-Fermi 
liquid system is that of interacting electrons in one dimension (1D) known as the Tomonaga-Luttinger model \cite{ REF:Tomonaga, 
REF:Luttinger}. By now, the main properties of this model are theoretically well-understood and form the concept of a
 \lq\lq Luttinger liquid\rq\rq \cite{ REF:Haldane81}. Luttinger liquids are very different from their higher dimensional 
counterparts, Fermi liquids, in many respects, including: The absence of single-particle excitations at low energies; spin-charge 
separation and the absence of a well-defined Fermi surface, even at zero temperatures. 

Despite the vigorous theoretical activity in this field, there has been only limited experimental evidence for the existence of 
Luttinger liquids in the conventional Q1D systems, such as organic conductors. The genuine 
Luttinger-liquid behavior is easily masked in these systems by other effects, such as Peierls transitions and  dimensionality 
crossovers  resulting from the coupling among the conducting chains. However, recent advances in semiconductor technologies have 
made  high-mobility quantum wires  new and promising candidates for studying  Luttinger-liquid effects in Q1D interacting 
electron systems. The most obvious
advantages of the quantum wires are: i)~the absence of unintentional dimensionality crossovers; ii)~a very low controllable 
degree of disorder and iii)~the absence of Peierls transitions.
Indeed, the first evidence for Luttinger-liquid behavior has recently been obtained in transport measurements \cite{ REF:Tarucha, 
REF:Tarucha2} on GaAs quantum wires, where the temperature dependence of the conductance of a weakly disordered single-channel 
\cite{ REF:Tarucha} and multi-channel \cite{ REF:Tarucha2} quantum wires was interpreted in terms of the Tomonaga-Luttinger model \cite{ REF:Yacoby}. 
Luttinger-liquid behavior has also been observed in transport  experiments on fractional quantum Hall systems \cite{ REF:Webb, REF:Chang}, 
where the edge states are believed to be in the chiral Luttinger-liquid state \cite{ REF:Wen}.

	Luttinger liquids and Fermi liquids are  two fixed-point regimes, which are well-understood on their own.
The crossover
between these regimes, which is expected to occur when several Luttinger liquids are
coupled together, is of  significant interest from the general theoretical point of view and
has been investigated by a number of authors 
\cite{ REF:Solyom, REF:Brazovskii, REF:Bourbonnais91, REF:Finkelstein93, REF:Nerseyan_2, REF:Fabrizio,  REF:Eduardo, 
REF:Bourbonnais95, REF:Balents, REF:Schulz96}.
However, it is difficult to compare the results of various theoretical approaches with the experiment, because the parameters driving this crossover such as the number of Luttinger-liquid
chains coupled together by  inter-chain tunneling, electron-electron interactions or both,
 cannot be changed smoothly in a real sample. The semiconductor 
quantum wires again appear to be ideal candidates for studying the Luttinger-liquid to Fermi-liquid  crossover,
because the number of conducting channels is an adjustable parameter of the experimental set-up. The indication for such 
a crossover in the conductance of GaAs quantum wires  has
recently been observed \cite{ REF:Tarucha2}. Theoretically, Matveev and Glazman \cite{ REF:Matveev_PRL, REF:Matveev} have used a multi-channel model to calculate tunneling into a clean wire.

	In this paper, we study the conductance of a multi-channel quantum wire in the presence
of disorder. Our main goal is to follow the crossover from the single-channel case to the multi-channel, when
the wire is expected to be in the Luttinger-liquid state and the Fermi-liquid state, respectively. The second motivation for this study comes
from  recent experiments \cite{ REF:Tarucha2}, in which, an indication
of such a crossover has been observed. Our main result is that the
temperature-dependent  conductance of a weakly disordered Luttinger-liquid wire, is reduced with an increasing number
of occupied channels $N$, and disappears in the limit of an infinite number of channels. We find that the scaling exponent
of the temperature dependence behaves as $1/N$, for $N\gg 1$. 

	This paper is organized as follows.  In Sec.~\ref{sec:model}, we describe the model
of a multi-channel quantum wire in the presence of long-range disorder and short-range electron-electron
interactions. In Sec.~\ref{sec:formalism}, we present a general formalism for the calculation of the
conductance and derive the expression for the exponent of the temperature scaling. This exponent
is analyzed for various situations  in Sec.~\ref{sec:exponent}. In Sec.~\ref{sec:Nspinless}, the general result for spinless
electrons is studied, and the comparison with the
experimental results is made in Sec.~\ref{sec:spincase}. Our conclusions are given in Sec.~\ref{sec:conclusions}.

\section{Formulation of the model}
\label{sec:model}
In this section, we outline the main assumptions and approximations used to calculate the temperature-dependent conductance of a weakly disordered Luttinger-liquid wire including: The geometry of the wire, the effects of disorder,
the nature of electron-electron interactions and the effect of electron reservoirs. We rely on the 
approaches developed by Glazman and Jonson \cite{ REF:Glazman} and by Matveev and Glazman \cite{ REF:Matveev}.
\subsection{Geometry}
\label{sec:geometry}
Consider a quantum wire of width $d$, {\it adiabatically} connected to the leads. For simplicity, the confinement
in the transverse direction is modeled by a square well-potential. The wavefunction
of the $n$th mode of transverse quantization  $\Psi_{n}(x,y)$ is expanded over the adiabatic basis of transverse wavefunctions
$\xi_{m\perp}(y)$
\begin{equation}
\Psi_{n}(x,y) = \sum_{m} \psi_{nm}(x) \xi_{m\perp}(y)\,.
\label{EQ:adiabat}
\end{equation}
Limiting to the leading (zeroth) order in the adiabatic expansion, $\psi_{nm}(x)$
takes the form
\begin{equation}
\psi_{nm}^{(0)}(x) =\frac{1}{\sqrt{L}} e^{ik_{n}x}\delta_{nm}\,,
\label{EQ:zeroth}
\end{equation}
where $k_{n}$ is the longitudinal wavevector of an electron with a total Fermi momentum $\hbar k_{F}$ 
\begin{equation}
k_{n} = k_{F} \sqrt{1 - \Big[\frac{n}{z}\Big]^{2}}\,,
\label{EQ:klong}
\end{equation}
where $z = \frac{k_{F} d}{\pi}$. The number of occupied transverse channels in the wire is
$N=[z]$. An {\em effective} Fermi velocity for channel $n$ is defined as $v_{F}(n) = \hbar k_{n} / m^{*}$.

\subsection{Disorder}
\label{sec:disorder}

In the absence of disorder, the conductance is quantized in units of $\frac{e^2}{h}$ per spin orientation, where each plateau
of quantization corresponds to a newly occupied channel.

The quasi-ballistic regime, where the wire length $L$ is shorter then the
(transport) mean free path $\ell$ is considered. This regime is realized in the experiments by Tarucha et al.
\cite{ REF:Tarucha, REF:Tarucha2}, in which $\ell/L>6$.

That the disorder potential in GaAs heterostructures varies slowly on
the scale of the Fermi-wavelength \cite{ REF:Ando}, {\it i.e.}, $k_F\ell_c>1$, where $\ell_c$ is the 
correlation length of the disorder potential \cite{FN:spacer} is also assumed. 
Backscattering processes in which the longitudinal momentum of the electron in the $i$th channel is changed by $\hbar\delta k_{ij}=\hbar k_i+\hbar k_j$, $j=1\dots N$, give the contribution to the resistance.
The probability of these processes $P_{BS}$ within a long-range disorder potential
 depends exponentially on $\delta k_{ij}$:
$P_{BS}\sim\exp(-2\delta k_{ij} \ell_c)$ \cite{ REF:Ando}. Therefore two different
regimes may be distinguished \cite{ REF:Glazman}. 

In the first regime,  $P_{BS}$ is exponentially
small for all occupied channels, except for the topmost one ($i=N$). In this channel, two situations can occur.
1) When the $N$ channel is just opened, the Fermi energy is equal to the threshold energy. The  momentum carried by this channel is small and is strongly affected by  impurity
scattering. In this case, the scattering in channel $N$ clearly gives the dominant contribution
to the resistance. 2) As the Fermi energy is increased, the momentum
increases, $P_{BS}$ decreases and finally becomes  exponentially small.
However, because $\delta k_{ij}$ is minimal for $i=j=N$, {\it i.e.}, for the backscattering within channel $N$,
this process dominates the resistance. Thus, regardless of the position of the Fermi energy to the threshold energy
the largest contribution to the resistance is given by the backscattering in the topmost channel , and the contributions from the 
rest of the channels are negligible. 

In the second regime,  $P_{BS}$ ceases to be exponentially
small for some channel $N_c<N$, such that $k_{N_c}\ell_c\sim 1$.  Then, all the channels with $N_c\leq n\leq N$
are subject to strong backscattering. 

As was shown by Glazman and Jonson \cite{ REF:Glazman}, the first (second) regime
is realized if $N<N^{*}$ ($N>N^{*}$), where $N^* =  8(k_F\ell_c)^2$.  In a typical
experimental situation, $N^{*}\simeq 100$ ($N^{*}\approx 60$ in the 
experiment \cite{ REF:Tarucha2}). As the number of observed plateau is usually significantly smaller than $N^*$, 
it suffices to consider only the first regime and take into account the backscattering only in the topmost
channel. 

Although the disorder is smooth on the scale of the Fermi wavelength ($k_F\ell_c >1$), it can be shown (cf. Appendix) that the effective potential
describing the backscattering of left- and right-moving excitations
can be represented by a $\delta$-function form: The only information on the smooth variations of the original
potential are hidden in the exponentially-small prefactor
of the $\delta$-function.

\subsection{Electron-electron interactions}
\label{sec:ee}

The same Luttinger-liquid model for a multi-channel wire as that proposed by Matveev and Glazman \cite{ REF:Matveev}is employed.
We specify the assumptions needed for this model and we begin with the case of spinless electrons.
The  Hamiltonian of interacting 2D spinless electrons  is given by:
\begin{equation}
H = H_{\circ} + H_{int}
\label{EQ:H}
\end{equation}
where $H_{\circ}$ is the Hamiltonian of free electrons and 
\begin{equation}
H_{int} = \frac{1}{2} \int d{\bf{r}} d{\bf{r'}} U({\bf r} - {\bf {r'}}) \hat{\Psi}^{\dagger}({\bf r}) \hat{\Psi}^{\dagger}({\bf {r'}}) \hat{\Psi}({\bf {r'}}) \hat{\Psi}({\bf r})\,,
\label{EQ:Ham}
\end{equation}
where $U({\bf r} - {\bf {r'}})$ is a (repulsive) Coulomb interaction and $\hat{\Psi}({\bf r})$ represents the fermionic 
field operator. 
Our first assumption is that the interaction term  (Eq.~\ref{EQ:Ham}) can be replaced by the direct density-density interaction
 between electrons occupying different channels, {\it i.e.},
\begin{equation}
H_{int} \Rightarrow \frac{1}{2} \sum_{ij} \int dx \int dx' \rho_i(x) \rho_j(x') U_{ij}(x - x')\,,
\label{EQ:Uij}
\end{equation}
where $\rho_k$ is the density operator of the $k-$th channel and 
\begin{equation}
U_{ij}(x - x') = \int dy'\!\int \!dy \,|\xi_i(y)|^2 |\xi_j(y')|^2 U({\bf r} - {\bf r}')\,.
\label{EQ:xixi}
\end{equation}
This assumption neglects the inter-channel exchange interactions, which are usually considered to be
less important than the direct ones due to the smaller values of the overlap integrals.

The Coulomb potential is assumed to be screened by the metallic gates forming the
channel, and in the {\it dc} limit the actual form of the potential $U({\bf r}-{\bf r}')$ can be replaced by the delta-function:
$U_{\circ }\delta({\bf r}-{\bf r}')$ \cite{ REF:Matveev}. 
For a $\delta$-function 2D potential, the effective 1D potential is also 
a $\delta$-function.
Using the eigenfunctions of a square-well confinement potential for $\xi_{i \perp}$ in Eq.~(\ref{EQ:xixi}), the 1D  
coupling constant is channel-independent and it is related to
$U_{\circ }$ by 
\begin{equation}
\hbar V_{\circ } = \frac{U_{\circ }}{d}\,,
\label{EQ:Vo}
\end{equation}
where the numerical coefficient has been absorbed into the redefinition of $U_{\circ }$.

The interaction Hamiltonian (Eq.~\ref{EQ:Uij}) causes forward and backward scattering processes.
In a multichannel case, the forward scattering is defined as the process in which none of the momenta of the electrons is reversed,
although the momentum transfer, $Q$, may not be equal to zero as the electrons can be exchanged between the channels.
Forward scattering includes processes with $Q\approx 0$ (for momentum transfer between electrons in the same channel and in different channels)
and $Q\approx k_F(i) - k_F(j)$ (for momentum transfer between electrons in channel $i$ and channel $j$).
The density-density interaction in Eq.~(\ref{EQ:Uij}) conserves the total number
of electrons in a given channel. Therefore, for temperatures low enough, {\it i.e.}, 
$T\ll\mbox{min}\{v_F(i), v_F(j)\}|k_F(i) - k_F(j)|$, 
the forward scattering with $Q \neq 0$ involves electron states only far away from the Fermi level
and can thus be neglected. 
Apart  from the renormalization of parameters, repulsive interactions the backscattering  do not change the low energy properties of the system \cite{ REF:Solyom, REF:Emery, REF:Schulz}, and therefore they are  not included
in the model \cite{ FN:backscattering}.
Finally, the Umklapp processes are not included,
because of the low  electron densities in semiconductor heterostructures: A typical quantum wire is very far away from the half-filling condition.

Each 1D channel is described by a Luttinger-liquid model in which the electron density fluctuations
of the $i$-th  channel are represented by a boson field $ \phi_{i}(x, \tau)$ defined so that
\begin{equation}
\rho_i(x) - {\bar\rho_i} = \frac{1}{\sqrt{\pi}} \partial_x \phi_i\,,
\label{EQ:density}
\end{equation}
where ${\bar\rho_i}$ is the average electron density in this channel.
The (number) current flowing in the $i$-th  channel is 
\begin{equation}
j = -i \partial_\tau \phi /\sqrt{\pi}\,.
\label{EQ:current}
\end{equation}
The (Euclidean) action of the system of interacting electrons occupying $N$ channels is given by \cite{ REF:Matveev}
\begin{mathletters}
\begin{eqnarray}
S_{1}& = & \frac{\hbar}{2}\!\int_{0}^{\beta} \!d\tau\!\int \!dx \sum_{i=1}^{N} \Big[\frac{1}{K_{i}v_{i}} (\partial_{\tau } \phi_{i})^{2} + \frac{v_{i}}{K_{i}} (\partial_{x} \phi_{i})^{2}\Big]\,,             
\label{EQ:Sone}                                       \\
S_{2}& = & \frac{\hbar}{2} \sum_{i\not= j}^{N} \frac{V_{\circ }}{\pi} \int_{0}^{\beta} d\tau dx \partial_x \phi_{i} \partial_x \phi_{j}\,.
\label{EQ:Stwo}
\end{eqnarray}
\end{mathletters}
The action $S_1$ describes  a set of $N$ Luttinger liquids with parameters $K_i, v_i$ which depend on the Fermi 
velocities $v_F(i)$ and the effective coupling constant $V_{\circ}$. The action $S_2$ describes the forward part of the
density-density interaction between  the channels. 
The electron spin will be included in 
Sec.~\ref{sec:spincase}.

\subsection{Effect of Reservoirs}
\label{sec:reservoirs}

Two characteristic features are predicted for the conductance of a single-channel Luttinger-liquid wire, $g$. First, in the absence of disorder, $g$ is expected to be
renormalized by the electron-electron interactions to the value of
$g=Ke^2/h$ per spin orientation \cite{ REF:Apel, REF:Kane and Fisher},
where $g=e^2/h$ for a non-interacting system, when $K=1$. Second, in the presence of weak disorder, $g$ had been shown to decrease with 
the temperature, revealing a tendency to interaction-enhanced Anderson localization \cite{ REF:Apel, REF:Giamarchi, 
REF:Ogata}. At temperatures lower than $T_L\equiv v_F/L$, this
temperature-dependence crosses over to
a length-dependence. However, 
as has recently been shown by a number
of authors \cite{ REF:Safi, REF:Dmitrii, REF:Ponomarenko, REF:Shimizu}, the first prediction does not survive if the presence
of the Fermi-liquid electron reservoirs attached to the wire,  is taken into account. Instead, the conductance remains
at its non-interacting value $g=e^2/h$. This result was obtained in Refs.~\cite{ REF:Safi, REF:Dmitrii, REF:Ponomarenko, REF:Shimizu}
in a model in which the Fermi-liquid reservoirs were imitated by switching off the interactions in the outer parts of the system,
{\it i.e.}, by putting $K=1$ outside the wire \cite{ FN:homogeneous system}. On the other hand, the second prediction 
was shown to survive even in the presence of the reservoirs 
\cite{ REF:DmitriiII, REF:SafiII, REF:Furusaki}. Moreover, the scaling exponent of the leading term in the
$T$-dependence was found to be independent of  the reservoirs, and the interaction strength behaved as uniform
throughout the system \cite{ REF:DmitriiII, REF:SafiII, REF:Furusaki, FN:correction by Safi and Schulz}.
This occurs because when $T\gg T_L$, {\it i.e.}, when $L\gg L_T$,
the density-density correlation function, whose $2k_F$ Fourier component determines the value of disorder-induced corrections
to the conductance, decays inside the wire and is only minimally affected by the presence of the reservoirs \cite{ FN:L-dependence}.
Thus, in order
to determine the temperature-dependence of the conductance,
the original model of a homogeneous Luttinger
liquid \cite{ REF:Apel, REF:Kane and Fisher, REF:Giamarchi, REF:Ogata} may be employed and 
the presence of the reservoirs may be ignored.
If  the interactions are not strongly attractive [\cite{ FN:correction by Safi and Schulz}], the error introduced by this simplification
 will be in an incorrect numerical prefactor of the
 $T$-dependence term, a non-universal quantity. Using this reasoning, we consider only the model of homogeneous coupled Luttinger liquids.

\section{CONDUCTANCE OF AN N-CHANNEL WIRE}
\label{sec:formalism}

In this section a general scheme for the calculation of the corrections to the conductance of a quantum wire carrying 
N occupied channels due to the presence of  weak disorder is presented. 

The current $I = ej$ is related to the electric field by
\begin{equation}
I(x,t) = \int_{-L/2}^{L/2} dx^{\prime } \int \frac{d\omega }{2\pi } e^{-i\omega t}\, \sigma_{\omega }(x,x^{\prime })\, E_{\omega }(x^{\prime })\,,
\label{EQ:current1}
\end{equation}
where $E_{\omega }(x)$ is the temporal Fourier component of the electric field and $\sigma_{\omega }(x,x^{\prime })$ is the non-local ac conductivity. To calculate $\sigma_{\omega}(x,x')$, we  make use of the Kubo formula \cite{ REF:Shankar90}
\begin{equation}
\sigma_{\omega }(x,x') = \frac{ie^2}{2\pi \hbar \omega } \int_{0}^{\beta } d\tau\, \langle T^{*}_{\tau } j(x,\tau )j(x',0)\rangle\, e^{i\bar{\omega }\tau }\, |_{\bar{\omega }=-i\omega +\epsilon}\,,
\label{EQ:Kubo}
\end{equation}
where $ej(x,\tau )$ is the total current through the wire, $T^{*}_{\tau }$ is a time ordered product as defined in 
Ref.~\cite{ REF:Shankar90} and $\bar{\omega }$ is the Matsubara frequency.

In the presence of $N$ channels, the total current is the sum of the currents carried by each channel. Upon bosonisation, 
the expression for the conductivity takes the form
\begin{equation}
\sigma_{\omega }(x,x^{\prime })   = \frac{e^2}{2\pi \hbar} \frac{i \bar{\omega }^{2}}{\omega } G_{\bar{\omega }} (x,x^{\prime }) |_{\bar{\omega} \rightarrow -i \omega + \epsilon}\,,             
\label{EQ:sigma}
\end{equation}
where 
\begin{equation}
G_{\bar{\omega }}(x,x^{\prime }) =  \int_{0}^{\beta } d\tau \sum_{i,j=1}^{N} \langle T_{\tau }^{*} \phi_{i} (x, \tau ) \phi_{j} (x^{\prime },0) \rangle e^{i \bar{\omega } \tau}\,,
\label{EQ:Greensf}
\end{equation}
and  $\phi_i$ is defined in Eq.~(\ref{EQ:density}).

For a 2D disorder potential $W({\bf r})$, the
effective 1D  potential $W_{ij}(x)$ is obtained by taking the matrix element of $W({\bf r})$ 
between the transverse wavefunctions $\xi_{i\perp}$ and $\xi_{j\perp}$. The action representing the impurity-electron 
interaction takes the form
\begin{equation}
S_I = \frac{\hbar}{2} \sum_{i,j}^N \int dx \, \psi_i^*(x)W_{ij}(x) \psi_j(x)\,,
\label{EQ:W2}
\end{equation}
where $\psi_k(x)$ is defined by Eq.~(\ref{EQ:zeroth}). Although the impurity scattering includes processes where electrons can be
transferred from channel $i$ to channel $j$, as discussed in Sec.~\ref{sec:disorder}, only the
backscattering  in the last occupied channel $N$ is important.
In this case,  $W_{ij}(x) = W(x) \delta_{iN} \delta_{jN}$.

The part of $S_I$ describing the backscattering is
\begin{equation}
S_{IB} = \frac{\hbar}{\pi a}\!\int_{0}^{\beta }\!d\tau\!\int \!dx\,W_B(x) \cos [2k_{F}(N) x + 2\sqrt{\pi } \phi_{N}]\,,
\label{EQ:Si}
\end{equation}
where $a$ is the microscopic cut-off length and $W_B(x)$ is the effective backscattering potential (cf. Appendix). 

We introduce the new fields $\chi (x,\tau)$ by rescaling the fields $\phi (x,\tau )$ as
\begin{equation}
\phi_{i}(x,\tau ) = \sqrt{K_{i}v_{i}}\, \chi_{i}(x,\tau)\,.
\label{EQ:xi}
\end{equation}
Using these fields, the action of the system without disorder takes the form
\begin{eqnarray}
S_1+S_2= \frac{\hbar}{2}\!\int_{0}^{\beta }\!d\tau \int \!dx \sum_{i,j=1}^{N} &\Big[&(\partial_{\tau} \chi_{i})^{2} + v_{i}^{2} (\partial_{x} \chi_{i})^{2} +                   \nonumber \\                           &     & V_{ij} (\partial_{x} \chi_{i}) (
\partial_{x} \chi_{j})\Big]\,,  
\label{EQ:Sxi}
\end{eqnarray}
where $V_{ij} = (V_{\circ }/\pi) \sqrt{K_{i}v_{i}K_{j}v_{j}}\,$.

Due to the separability of the interaction term (Eq.~\ref{EQ:Stwo}), this action can be diagonalized exactly to give
\begin{equation}
S_1+S_2 = \frac{\hbar}{2} \int_{0}^{\beta } \!d\tau \int \!dx \sum_{i=1}^{N} \Big[(\partial_{\tau} \tilde{\chi}_i)^{2} + w_{i}^{2} (\partial_{x} \tilde{\chi}_i)^{2}]\,,
\label{EQ:Sdiagonal}
\end{equation}
where $\tilde{\chi}_m = \sum_{i=1}^N A_{mn} \chi_n$ and $w_{i}$ are the eigenvalues of $A_{mn}$  satisfying the following equation:
\begin{equation}
\frac{\pi}{V_{\circ }} = \sum_{j=1}^{N} \frac{K_{j}v_{j}}{(w_i^{2} - v_{j}^{2} + (V_{\circ }/\pi ) K_{j}v_{j})}\,. 
\label{EQ:eigenva}
\end{equation}
The elements of the diagonalization matrix $A$ are given by
\begin{eqnarray}
A_{ij}^{2} &=& \frac{K_{i}v_{i}}{(w_{j}^{2} - v_{i}^{2} + (V_{\circ }/\pi ) K_{i}v_{i})^{2}} \nonumber    \\
           & & \Big[\sum_{l=1}^{N} \frac{K_{l}v_{l}}{(w_{j}^{2} - v_{l}^{2} + (V_{\circ }/\pi ) K_{l}v_{l})^{2}} \Big]^{-1}\,.
\label{EQ:eigenve}
\end{eqnarray}
The expression for the impurity action now takes the  form
\begin{eqnarray}
S_{I} = \frac{\hbar}{\pi a}\int_{0}^{\beta } &d\tau& \int dx \,W(x) \cos [2k_{F}(N) x + \nonumber   \\
                                             &     &\sqrt{4 \pi } \sum_{i=1}^{N} A_{Ni}\sqrt{K_{i}v_{i}} \tilde{\chi}_i]\,.
\label{EQ:Sixi}
\end{eqnarray}
Because of the assumption of weak disorder, the conductance $g$ can
be obtained via the perturbation expansion in $W(x)$:
\begin{equation}
g = N \frac{e^2}{2\pi \hbar} + \delta g\,.
\label{EQ:g}
\end{equation}
The leading term in Eq.~(\ref{EQ:g}) is taken to be unrenormalized by the interactions (cf. Sec.~\ref{sec:disorder}). 
$\delta g$ is expressed in terms of the correlation function $\overline{W(x)W(0)}$.
As is shown in the Appendix, this correlation function  can be taken in the following
form
\begin{equation}
\overline{W_B(x_{1})W_B(x_{2})} = n_{i}u^{2} \delta(x_{1}-x_{2}) \,,
\label{EQ:impcorr}
\end{equation}
where $u$ is the effective impurity strength. 
The only information about the long-range nature of the disorder is now contained in the parameter $u^2$, which is proportional
to the backscattering probability $P_{BS}$ of the original potential (cf.~Sec.\ref{sec:disorder}).
Under this assumption, the leading (second order in W) contribution for the correction to the Green's function  is given by
\begin{eqnarray}
& &\delta G_{\bar{\omega}}(x,x^{\prime})= -\frac{2 n_{i} u^{2}}{(\pi a^{2})} \int_{-L/2}^{L/2} dx_{1} [F_{0}^{(N)}(x_{1}) - F_{\bar{\omega}}^{(N)}(x_{1})] \nonumber \\
& &\Big[\sum_{i,j,l,m=1}^{N} c_{il}c_{jm}c_{Nl}c_{Nm} \tilde{G}_l(\bar{\omega};x-x_{1}) \tilde{G}_m(\bar{\omega};x_{1}-x^{\prime})\Big]
\label{EQ:deltag1}
\end{eqnarray}
where 
\begin{equation}
F_{\bar{\omega}}^{(N)}(x) \! = \!\int_{0}^{\beta} \! d\tau e^{i \bar{\omega} \tau}\exp(-4\pi \sum_{j=1}^{N} c_{Nj}^{2} [\tilde{G}_j(x,0) - \tilde{G}_j(x,\tau)])\,,
\label{EQ:2kfcorrel}
\end{equation}
$\tilde{G}_j(x, \tau)$ is the propagator of the field $\tilde{\chi }_j$ defined as
\begin{equation}
\tilde{G}_j(x-x', \tau) = -\langle T_{\tau} \tilde{\chi}_j(x, \tau) \tilde{\chi}_j(x', 0) \rangle\,,
\label{EQ:Gchi}
\end{equation}
and  $c_{ij} = \sqrt{K_{i}v_{i}} A_{ij}$. The function $F_{\bar{\omega}}^{(N)}$ is the Fourier transform of the $2k_{F}$ density-density
correlation function for the channel $N$.
Because the action $S_1+S_2$ as written in Eq.~(\ref{EQ:Sdiagonal}) is quadratic in $\tilde{\chi}_j$,
the propagator $\tilde{G}_j$ and the
function $F_{\bar{\omega}}^{(N)}$ can be calculated  straightforwardly. Following the procedure of Ref.~\cite{ REF:DmitriiII}, 
we obtain for $L \gg L_T$ 
that the correction to the conductance is given by
\begin{equation}
\delta g = - \frac{e^2}{2\pi \hbar} C \frac{L}{\ell^*} \Big(\frac{4\pi T}{\hbar \omega_{F}}\Big)^{-\alpha_N}
\label{EQ:deltag2}
\end{equation}
where
\begin{equation}
\alpha_N = 2 \Big(1 - K_{N}v_{N}\sum_{j=1}^{N} \frac{A_{Nj}^{2}}{w_{j}}\Big)\,.
\label{EQ:alpha}
\end{equation}
In (\ref{EQ:deltag2}), $\ell^*$ is an effective elastic mean free path  defined 
by $1/ \ell^* = n_{i}u^{2}/a^{2}\omega _{F}^{2}$~, with $\omega_{F}$ being the (non-universal) ultraviolet 
energy cut-off, and $C$ is a positive constant depending on $K_i, v_i$. Eq.~(\ref{EQ:deltag2}) is the most general result given in 
terms of the initial parameters of the model. 

In the next section various regimes in which  the scaling  exponent
$\alpha $ can be calculated are discussed.

\section{SCALING EXPONENT}
\label{sec:exponent}

\subsection{One and N channels for spinless electrons}
\label{sec:Nspinless}

The case of a wire with only one occupied spinless channel is the simplest one and the well-known value for the temperature exponent
is easily recovered. In this case $N = 1$, $A_{11} = 1$, and from the eigenvalue 
equation (\ref{EQ:eigenva}) $w = v$. Thus, the temperature exponent reduces to $\alpha_1 = 2(1 - K)$ in
agreement with  previous results \cite{ REF:Apel, REF:Giamarchi}.

To analyze the case of  $N$ occupied channels, it is necessary to solve the eigenvalue equation (\ref{EQ:eigenva})
which amounts to finding all zeroes of an $N$ degree polynomial.
The solutions to this equation ($w_i $) have the meaning of the sound velocities of the new fields 
${\tilde{\chi}_i}$, whereas the original fields $\phi_i $  propagate with velocities $v_i$. 
The  main feature of 
these solutions  is that all but one $w_i $ lie
between the values of two adjacent $v_i$ ($v_{i-1} < w_i < v_i$), whereas one $w_j$ is bigger than the maximal $v_i$. This 
biggest velocity  corresponds to the collective mode ${\tilde{\chi}_j} = \sum_i c_{ji} \phi_i$ where all the coefficients $c_{ji}$ 
have the same sign. The linear combination for the rest of $w_i$ include coefficients $c_{il}$ with different relative signs.

In a model case where the $N$ coupled channels are viewed as $N$ equivalent coupled chains with disorder scattering only in one chain, the eigenvalue equation for $N \geq 2$ can be solved exactly. In this case
$K_i= K$; $v_i = v$, $i = 1,\ldots , N$; Eq.~(\ref{EQ:eigenva}) can be solved analytically and the matrix elements $A_{ij}$ can be found explicitly. The 
expression for the temperature exponent is given by
\begin{eqnarray}
\alpha_N & = & 2\Big[1 - \frac{1}{N} \frac{K}{\sqrt{1 + [V_{\circ } K (N-1)]/(\pi v)}} -  \nonumber  \\
       &   & (1 - \frac{1}{N}) \frac{K}{\sqrt{1 - (V_{\circ }K)/(\pi v)}}\Big]\,. 
\label{EQ:alphaN}
\end{eqnarray}

In Eq.~(\ref{EQ:alphaN}), the second term is the contribution coming from the collective mode, and the last one 
is the contribution from all the other modes.
In the limit of large $N$, we obtain a finite value for $\alpha$, for generic values of $K$, $v$ and $V_{\circ }$. 
However, these parameters are not independent, but are related as 
\begin{mathletters}
\begin{eqnarray}
K & = & \Big[1 + \frac{V_{\circ }}{\pi v_F}\Big]^{-1/2}\,,
\label{EQ:K}             \\
v & = & v_F\Big[1 + \frac{V_{\circ }}{\pi v_F}\Big]^{1/2}\,.
\label{EQ:v}
\end{eqnarray}
\end{mathletters}
Substituting these equations into Eq~(\ref{EQ:alphaN}), we find:
\begin{equation}
\alpha_N = \frac{2}{N}(1 -\Big[1 + \frac{V_{\circ }N}{\pi v_{F}} \Big]^{-1/2}) 
\label{EQ:alphab}\,.
\end{equation}
In the limit
$N \rightarrow \infty$, the  exponent vanishes rendering a temperature independent conductance. 

In order to understand the temperature dependence of the conductance for quantum wires, however,  we need to work with a set of $2N$ 
different parameters $K_i, v_i$. These parameters are related to the Fermi 
velocity of the channel by $K_iv_i = v_F(i)$, with $v_F(i)$ defined in Sec.~\ref{sec:geometry}.
Making  use of relations similar to those
given by Eqs.(\ref{EQ:K},\ref{EQ:v})
\begin{mathletters}
\begin{eqnarray}
K_i & = & \Big[1 + \frac{V_{\circ }}{\pi v_F(i)}\Big]^{-1/2}\,,
\label{EQ:Ki}             \\
v_i & = & v_F(i)\Big[1 + \frac{V_{\circ }}{\pi v_F(i)}\Big]^{1/2}\,,
\label{EQ:vi}
\end{eqnarray}
\end{mathletters}
the eigenvalue equation reduces to
\begin{equation}
\frac{\pi }{v_{\circ}} = \sum_{i=1}^N \frac{s_i}{\tilde{w}_j^2 - s_i^2}
\label{EQ:eigenva2}
\end{equation}
where $v_{\circ} = V_{\circ }/v_F$,  $s_j = \sqrt{1 - (j/z)^2}$ and $\tilde{w}_j = w_j/v_F$. 
Under the conditions $N \gg 1, V_{\circ }/v_F \ll 1$, and $V_{\circ } N \rightarrow$ const,
\begin{equation}
\alpha_N = \frac{2}{N} \frac{V_{\circ }N}{\pi v_F} [1 -{\cal O} (\frac{1}{N})]\,.
\label{EQ:alphac}
\end{equation}

Eq.~(\ref{EQ:alphac}) contains the combination 
$V_{\circ } N /v_{F}$. Using Eq.~(\ref{EQ:Vo}) 
in the limits $N \rightarrow \infty$ and $V_{\circ }\rightarrow 0$, this combination approaches the constant value $U_{\circ }m/\pi \hbar^2$, 
which is the 
dimensionless coupling constant of the original 2D problem \cite{ REF:Eduardo}, common in  Fermi-liquid
theory \cite{ REF:Baym}. 
Thus, in both cases of $N$ occupied channels and $N$ chains,  the dependence of the temperature
exponent with the number of channels (or chains) is $1/N$ as shown in Eqs.~(\ref{EQ:alphab}, \ref{EQ:alphac}). This result
is in agreement with the dependence found by Matveev and Glazman \cite{ REF:Matveev} for the exponent of the tunneling conductance,
after redefining  the value of their one-dimensional interaction potential in such a way to get a finite two-dimensional 
coupling constant. In the same limit, the velocity of the collective mode $\tilde{\chi}_j$ approaches the zero-sound velocity of a two-dimensional
Fermi-liquid \cite{ REF:Baym, REF:Eduardo}.

As is seen from Eqs.~(\ref{EQ:alphab}) and (\ref{EQ:alphac}), in the limit $N \rightarrow \infty$
the temperature exponent vanishes, and the correction to  the conductance , $\delta g$, becomes temperature 
independent. This corresponds to the conductance of a Fermi-liquid at low temperatures in the presence of weak disorder.
In the Born approximation, lowest order of perturbation theory in impurity scattering,
the temperature-dependent weak-localization corrections are not observed.
Thus, as the number
of occupied channels (or the number of chains) increases, the crossover between a Luttinger-liquid  and a 2D Fermi-liquid is observed.

\subsection{N = 2 (Electrons with spin)}
\label{sec:spincase}

Finally, in order to compare  theoretical with the experimental results \cite{ REF:Tarucha2}, the temperature exponent 
for a two-channel wire for the case of electrons with spin is calculated. To include spin, a boson field for each spin 
orientation $\phi_{\sigma , i }$ ($\sigma = \uparrow, \downarrow$) is introduced, 
and  the charge and spin fields are defined as follows
\begin{mathletters}
\begin{eqnarray}
\phi_{c,i}& = & \frac{\phi_{\uparrow ,i} + \phi_{\downarrow ,i}}{\sqrt{2}}\,,
\label{EQ:phic}    \\
\phi_{s,i}& = & \frac{\phi_{\uparrow ,i} - \phi_{\downarrow ,i}}{\sqrt{2}}\,.
\label{EQ:phis}
\end{eqnarray}
\end{mathletters}
The action in the absence of disorder is 
\begin{eqnarray}
{\tilde{S}} = \frac{\hbar }{2} \int_0^{\beta }& & d\tau \int dx \Big\{\sum_{\mu =c,s} \sum_{j=1}^2 \Big[\frac{1}{K_{\mu j}v_{\mu j}} (\partial_{\tau } 
\phi_{\mu j})^2              \nonumber \\
                                   +          & & \frac{v_{\mu j}}{K_{\mu j}} (\partial_x \phi_{\mu j})^2 \Big] + (V_{\circ }/\pi ) \partial_x \phi_{1c} \partial_x \phi_{2c} \Big\}\,,
\label{EQ:spinaction}
\end{eqnarray}
where the parameters $K_{\mu j}, v_{\mu j}$ correspond to the charge and spin Luttinger-liquid parameters \cite{ REF:Schulz} for 
both channels. The backscattering part of the interaction has not been included in  Eq.~(\ref{EQ:spinaction})  
according to the discussion in Sec.~\ref{sec:ee}.
The part of the impurity action describing the backscattering in the topmost ($i = 2$) channel has the form
\begin{eqnarray}
{\tilde{S}}_{IB} =\!\frac{2\hbar }{\pi a}\!\int_0^{\beta}\!\!d\tau\!\int\!dx\, &W(x) & \cos[2k_F(2)x + \sqrt{4\pi } \phi_{2c}(x,\tau )] \nonumber \\
                                                                               &\cos(&\sqrt{4\pi } \phi_{2s}(x,\tau))\,.
\label{EQ:spinimpur}
\end{eqnarray}
Following the same procedure as in the spinless case, the temperature exponent of the conductance is
\begin{equation}
\tilde{\alpha}_2 = 1 - K_{2c}v_{2c}\sum_{i=1}^2 \frac{A_{2ic}^2}{w_{ic}}\,,
\label{EQ:alphad}\,
\end{equation}
where $A_{lmc}$ is the matrix used to diagonalize the charge part of the action and $w_{ic}$ are the corresponding eigenvalues.
Here, the $SU(2)$ symmetry requirement of $K_{2s} = 1$ has already been  satisfied.
In complete analogy with the spinless case, the relation between $K_{ci}, v_{ci}$ and the Fermi velocity of a
given channel and the equality between the intra- and inter-channel interactions are used to solve the eigenvalue equation. 
For completeness, explicit analytic expressions for the eigenvalues and matrix elements are presented
\begin{mathletters}
\begin{eqnarray}
s_{1,2}^2 & = &\frac{x_1^2 + x_2^2}{2} + \frac{g}{2}(x_1 + x_2) \pm \nonumber \\
          &   &\frac{(x_1^2 - x_2^2)}{2} \sqrt{1 + \frac{2g}{x_1+x_2} + \frac{g^2}{(x_1 - x_2)^2}} 
\label{EQ:s12}\, ;\\
A_{2jc}^2 &  =&\frac{1}{\Big[1 + \frac{x_1}{x_2}\Big(\frac{s_j^2 - x_2^2}{s_j^2 - x_1^2}\Big)^2\Big]}\,  ,
\label{EQ:a2jc}
\end{eqnarray}
\end{mathletters}
where $x_i =\sqrt{1 - (i/z)^2}$, $s_{i} = w_{ic} / v_F$ and $g = V_{\circ }/(\pi v_F)$.
In the one-channel case with spin,
\begin{equation}
\tilde{\alpha}_1 = 1 - K_{c1}\,.
\label{EQ:alphae}
\end{equation}
   
In order to compare the one- and two-channel cases, the exponents $\tilde{\alpha}_1$ and $\tilde{\alpha}_2$ are plotted
as a function of $K_{c1}$ as an effective measure of the interaction strength.
As is shown in Fig.~\ref{FIG:fig}, the value
of the exponent for two channels is smaller than for one channel. This is in  qualitative 
agreement with the experimental observations \cite{ REF:Tarucha2}, where the temperature dependence of the conducting wire was
found to be  weaker than in the one-channel case. The arrow indicates the value of $K_{c1} \simeq 0.7$ obtained from the
analysis of the data on the conductance of a single-channel wire \cite{ REF:Tarucha2}. For this value of $K_{c1}$, 
$\tilde{\alpha}_2$ is smaller than $\tilde{\alpha}_1$ by only about $30\%$. Thus, although the Fermi-liquid-like behavior sets in 
for $N \gg 1$,  a Luttinger-liquid behavior is still well-pronounced
for a few-channel wire.
\section{CONCLUSIONS}
\label{sec:conclusions}
The correction to the conductance of an $N$ channel quantum wire arising from weak impurity scattering is calculated, assuming that
the wire can be modeled by a set of $N$ coupled Luttinger liquids.

The general expression for the scaling exponent
of the temperature dependent conductance of a quantum wire is obtained, and  it is shown that this exponent behaves as $1/N$ for $N\gg 1$. That is,  as the number of channels increases, the temperature dependence diminishes and vanishes in the limit $N \rightarrow \infty$.
This temperature-independent conductance is  characteristic for a  Fermi-liquid system.
In this way, the results presented support 
the idea of a continuous crossover between the Luttinger-liquid and the Fermi-liquid pictures, as the number of channels (or chains)
is increased, as has been suggested by several authors \cite{ REF:Solyom, REF:Eduardo}.

Finally, the temperature exponent in the two-channel case is calculated including the electron spin. The result agrees 
qualitatively with the experimental observations.

\section{acknowledgments}
We thank Eduardo Fradkin for his continuous interest in this work  and many important suggestions and Seigo Tarucha and Yasuhiro Tokura for the valuable discussions
of the experimental results. This work was supported by the Department of Physics
of the University of Illinois at Urbana-Champaign (NPS), NSF under
grant DMR-89-20538, and the Department of Physics, University
of Florida at Gainesville (DLM). The work of NPS was also supported in
 part by  an International Fellowship from AAUW. DLM gratefully acknowledges the hospitality the NTT
Basic Research Laboratories, where part of this work has been done.
\vspace{1cm}

After this work was completed, we have learned of a recent preprint by Kawabata and Brandes, who have also found a $1/N$ scaling
for the temperature exponent by using a different method.

\appendix{}

In this Appendix, the bosonized expression
 for the impurity backscattering Hamiltonian  for the one-channel case is derived. 
This calculation is necessary because, as will be shown
below, the effective backscattering potential
for the right- and left-movers is different from the original
one. It is shown that even if the original potential
is long-ranged, the effective backscattering potential
takes the local ($\delta$-function) form. 

For a one-channel quantum wire, the Hamiltonian describing 
the electron-impurity interaction is
\begin{equation}
H = \int dx \psi^{\dag}(x) W(x) \psi(x) 
\label{EQ:apW2}
\end{equation}
or, in momentum space,
\begin{equation}
H = \int_{-\infty}^{\infty} \frac{dq}{2\pi} \frac{dQ}{2\pi} \psi^{\dag}(q+Q) \psi(q) W(Q)\,,
\label{EQ:H(q)}
\end{equation}
where $W(x)$ is the disorder potential, whose correlation function
$\overline{W(x)W(0)}$ is assumed to be known.
The operators of right- and left- moving electrons are introduced 
and the contributions to the integral over $Q$ in Eq.~(\ref{EQ:H(q)})
from the forward ($|Q| \approx 0$) and backward 
($|Q| \approx 2k_F$) scattering processes are separated. Going back to the real-space representation, the backscattering part of Hamiltonian (\ref{EQ:apW2}) takes the form
\begin{equation}
H_B = \int \frac{dx}{\pi a} \cos(2k_Fx + \sqrt{4\pi} \phi) W_B(x)\, ,
\label{EQ:Hbcos}
\end{equation}                                      
where $W_B(x)$ is the effective backscattering potential
\begin{equation}
W_B(x) = 2Re\int_{2k_F-\Lambda}^{2k_F+\Lambda} \frac{dQ}{2\pi} \,\,W(Q) \,\,e^{iQx}\,,
\label{EQ:W2kf}
\end{equation}
and $\Lambda = 1/a \ll k_F$ is a {\it hard}  momentum cutoff.
Note that $W_B(x)$ is not equal to the original $W(x)$ because
the integration over $Q$ in Eq.~(\ref{EQ:W2kf}) is taken over  narrow
regions near $2k_F$ only. Representing $W(x)$ by a sum of single impurity potentials
$u(x)$, the correlation function $\overline{W_B(x)W_B(0)}$ takes the form
\begin{equation}
\overline{W_B(x)W_B(0)}= 2 n_i \int_{2k_F-\Lambda}^{2k_F+\Lambda} \frac{dQ}{2\pi}
|u(Q)|^2 \cos Qx \, .
\label{EQ:e-2kfxi}
\end{equation}
To avoid spurious oscillations introduced by a hard cutoff procedure, an actual calculation
should be performed by using the soft cutoff procedure. Therefore,
the following change is made
\begin{equation}
\int_{2k_F-\Lambda}^{2k_F+\Lambda} \frac{dQ}{2\pi}\dots\Rightarrow
\int_{-\infty}^{\infty} \frac{dQ}{2\pi}e^{-\frac{(Q-2k_F)^2}{2\Lambda^2}}
\dots \, .
\label{EQ:soft}
\end{equation}
$u(Q)$ is chosen to be
\begin{equation}
u(Q) = u_{\circ} e^{-|Q|\ell_c}\, .
\label{EQ:uq}
\end{equation}
This form captures correctly the exponential dependence of $u(Q)$ on $Q$ 
for a realistic disorder potential in $GaAs$ heterostructures 
\cite{ REF:Ando}. Performing the integration in Eq.~(\ref{EQ:e-2kfxi}), one obtains
\begin{eqnarray}
\overline{W_B(x)W_B(0)}= n_i\big(u_{\circ} e^{-2k_F\ell_c}\big)^2\times
\nonumber\\
\big[g_{1\Lambda}(x)
\cos(2k_Fx)+
g_{2\Lambda}(x)\sin(2k_Fx)\big]\, ,
\label{EQ:wbcorr}
\end{eqnarray}
where
\begin{equation}
g_{1,2\Lambda}(x)=\Lambda \sqrt{\frac{2}{\pi}}e^{2\Lambda_2\ell_c^2}e^{-\frac{\Lambda^2x^2}{2}}\times
\left\{
\begin{array}{l}
\cos(2\Lambda^2\ell_c x)\\
\sin(2\Lambda^2\ell_c x)\\
\end{array}\right .\, .
\end{equation}

The correction to the bosonic propagator is given by\\
\begin{mathletters}
\begin{eqnarray}
&&\delta G_{\bar{\omega}}(x-x')=-\frac{2}{\pi a^2} \Big(\frac{2\pi}{\omega_F \beta}\Big)^{2K_1} \int_{-L/2}^{L/2} dx_1dx_2 \nonumber \\
&&\overline{W_B(x_1-x_2)W_B(0)} \cos 2k_F(x_1-x_2) \nonumber \\  
&& \Big[G^{0}_{\bar{\omega}}(x-x_1) G^{0}_{\bar{\omega}}(x_1-x') F_{0}(x_1-x_2)
 - \nonumber \\
&& G^{0}_{\bar{\omega}}(x-x_1) G^{0}_{\bar{\omega}}(x_2-x') F_{\bar{\omega}}(x_1-x_2)\Big]\,
;\label{EQ:Gcorr}\\
&&F_{\bar{\omega}}(x)=\int_0^{\beta} d\tau \frac{e^{i\bar{\omega} \tau}}
{[(\sinh \pi x/L_T)^2 + (\sin \pi \tau /\beta)^2]^{K_1}}\,,
\label{EQ:dens-dens}
\end{eqnarray}
\end{mathletters}
where $G^{0}_{\bar{\omega}}(x)$ is the propagator in the absence of disorder.
The product $\overline{W_B(x)W_B(0)} \cos 2k_Fx$
can be separated as
\begin{eqnarray}
                 & &\overline{W_B(x)W_B(0)} \cos 2k_Fx\propto\nonumber\\
g_{1\Lambda}(x)+ & & \cos(4k_Fx)g_{1\Lambda}(x)+\sin(4k_Fx)g_{2\Lambda}(x)\, .
\label{EQ:prod}
\end{eqnarray}
An estimate of the length-scales of the various
functions entering the integrals over $x_{1,2}$ in Eqs.~(\ref{EQ:Gcorr},\ref{EQ:dens-dens}) can be done. First of all, the product $\overline{W_B(x)W_B(0)} \cos 2k_Fx$
contains a component oscillating on the scale $\simeq 1/4k_F$, whereas the functions
$g_{1,2\Lambda}(x)$ in this product oscillate on the 
scale $\simeq 1/(2\Lambda^2\ell_c)$ and decay rapidly on the scale
$\simeq 1/\Lambda$. The density-density correlation
function (\ref{EQ:dens-dens}) decays on the scale $\simeq L_T$.
The propagator $G^{0}_{\bar{\omega}}(x)$ decays on the scale
$\simeq v_F/\bar{\omega}$. After the analytic continuation
($i\bar{\omega}\rightarrow \omega+i0$) is performed and the
the {\it dc} limit is taken, this scale becomes infinite.
By the meaning of the cutoff procedure, $L_T,\ell_c\gg1/\Lambda$.
Also, when comparing the scales of $1/4k_F$ and $1/2\Lambda^2\ell_c^2$,
one has to recognize that the limit $k_F/\Lambda\rightarrow\infty$ is to be taken
before the limit $\Lambda\rightarrow\infty$. 
Thus it is posible to establish  the following hierarchy of scales
\begin{equation}
\frac{1}{4k_F}\ll\frac{1}{\Lambda} \left(\frac{1}{2\Lambda \ell_c}\right)\ll \frac{1}{\Lambda}\ll L_T\,.
\label{EQ:ineqleng}
\end{equation}
This shows that the $4k_F$-oscillating terms in Eq.~(\ref{EQ:prod}) can be neglected
as these oscillations are the most rapid ones, whereas the
function $F_{\bar{\omega}}(x)$ varies slowly compared to
$g_{1\Lambda}(x)$.  As can be easily checked,
$g_{1\Lambda(x)}|_{\Lambda\to\infty}\to\delta(x)$. The effective correlation
function takes the form
\begin{equation}
\overline{W_B(x)W_B(0)}\to n_i\big(u_{\circ} e^{-2k_F\ell_c}\big)^2\delta(x)\, ,
\label{EQ:effcorr}
\end{equation}
which is the same as for a sum of $\delta$-function impurities with the exception that 
the strength of each impurity is renormalized. This renormalization is
the only effect of the actual form of the impurity potential on the
effective backscattering potential. For a $\delta$-function original
potential ($\ell_c=0$), the renormalization is absent. For a long-ranged
potential ($k_F\ell_c >1$), such as the present in $GaAs$ heterostructures,
the backscattering potential is exponentially weak.


\begin{references}
\bibitem{FN:spacer} $\ell_c$ is typically given by the distance between the layer
of a two-dimensional electron gas and the plane of ionized donors $z_0$\cite{ REF:Ando}, which is usually
of the order of $300-700\AA$.
\bibitem{ FN:backscattering} Strictly speaking, in a spinless model the backscattering is simply an exchange process to the forward
scattering, and in this sense both  interactions within a given channel are included.
\bibitem{ FN:correction by Safi and Schulz} Safi and Schulz \cite{ REF:SafiII} have recently found the sub-leading
terms in the $T$-dependence which come from the scattering off impurities located near the interfaces between
the wire and the reservoirs. These terms becomes dominant however, only if the interaction is strongly attractive
[$K>(3+\sqrt{17})/4$], which is not a relevant case for semiconductor quantum wires.
\bibitem{ FN:L-dependence} The exponent of the $L$-scaling, which replaces the $T$-scaling for $T\ll T_L$,
also remains the same as if the reservoirs were absent \cite{ REF:DmitriiII}. The physical reason for this is apparently
different however from that given above.
\end{references}
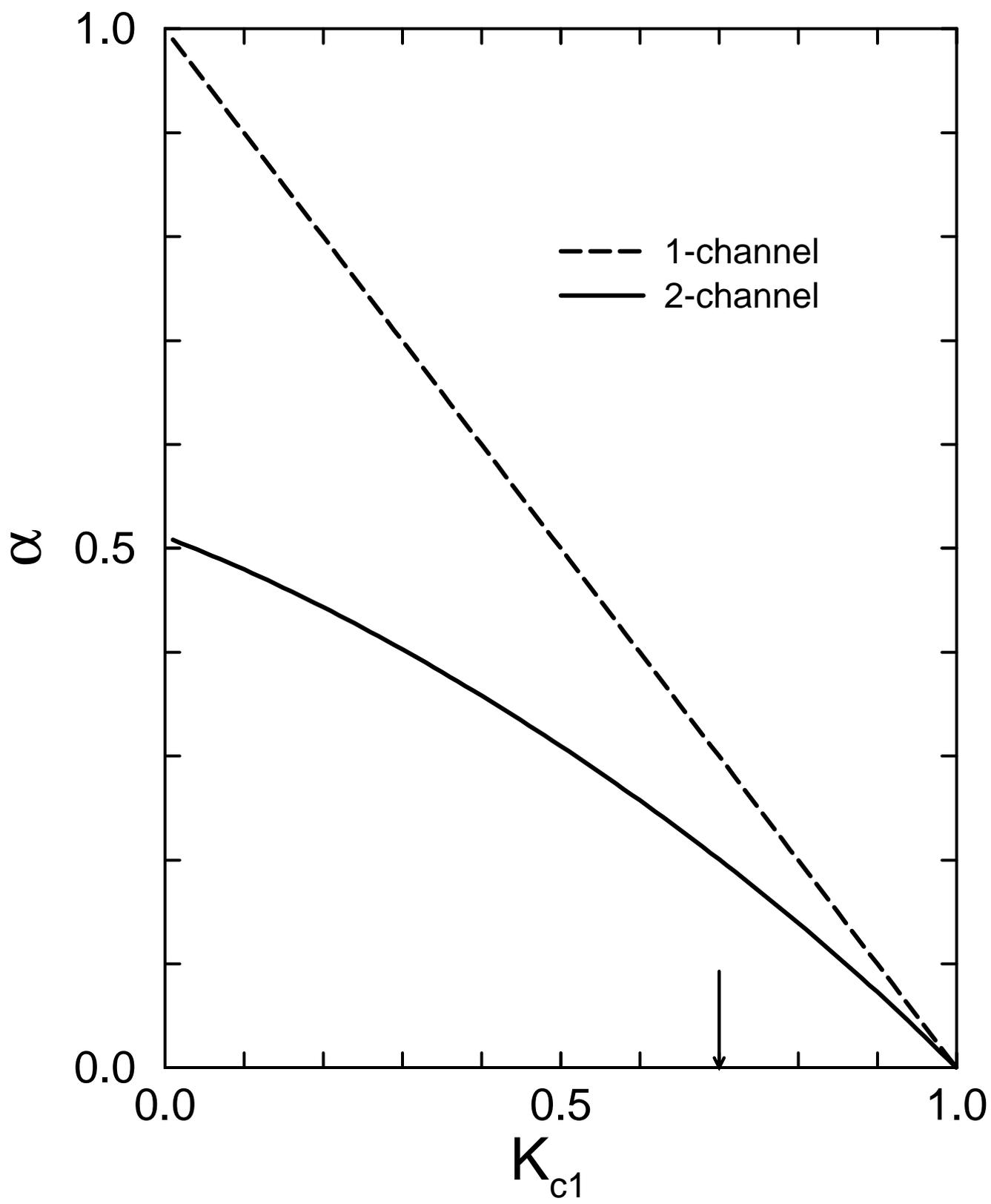
\end{document}